\documentclass[twocolumn,showpacs,preprintnumbers,amsmath,amssymb, showkeys]{revtex4}
\usepackage{graphicx}% Include figure files
\usepackage{dcolumn}% Align table columns on decimal point
\usepackage{bm}% bold math
\usepackage{epsf}

\begin{document}

%\preprint{APS/123-QED}

\title{Spatial distribution of local tunneling conductivity due to interference and Coulomb interaction effects for deep and shallow impurities on semiconductor surfaces}

\author{V.\,N.\,Mantsevich}
 \altaffiliation{vmantsev@spmlab.phys.msu.ru}
\author{N.\,S.\,Maslova}%
 \email{spm@spmlab.phys.msu.ru}
\affiliation{%
 Moscow State University, Department of  Physics,
119991 Moscow, Russia
}%

\date{\today }
5 pages, 3 figures
\begin{abstract}
Spatial distribution of local tunneling conductivity was
investigated for deep and shallow impurities on semiconductor
surfaces. Non-equilibrium Coulomb interaction and interference
effects were taken into account and analyzed theoretically with the
help of Keldysh formalism. Two models were investigated: mean field
self-consistent approach for shallow impurity state and Hubbard-{I}
model for deep impurity state. We have found that not only above the
impurity but also at the distances comparable to the lattice period
both effects interference between direct and resonant tunneling
channels and on-site Coulomb repulsion of localized electrons
strongly modifies form of tunneling conductivity measured by the
scanning tunneling microscopy/spectroscopy (STM/STS).
\end{abstract}

\pacs{71.55.-i}
\keywords{D. Fano resonance; D. Coulomb interaction; D. Mixed valence; D. Hubbard {I}; D. Impurity; D. Tunneling conductivity spatial distribution}%Use showkeys class option if keyword
                              %display desired
\maketitle

\section{Introduction}

    Impurity states at surface and interfaces of semiconductors strongly
modify the local electronic structure and consequently determine the
behavior of tunneling characteristics in STM/STS contacts
\cite{Dombrowski,Sullivan,Mahieu}. Experimental and theoretical
investigations of tunneling through impurity atom energy level in
the case of multichannel transport reveal Fano-type line shape in
local tunneling conductivity  when STM metallic tip is positioned
above the impurity \cite{Fano,Gores,Boon}. Transformation of
Fano-type line shape in local tunneling conductivity depending on
distance value from impurity was also recently investigated \cite
{Madhavan,Mantsevich}. Most of the experiments are carried out with
the help of scanning tunneling microscopy/spectroscopy technique
\cite {Dombrowski,Boon,Maslova,Panov} and theoretical calculations
usually deals with Green's functions formalism
\cite{Mantsevich,Hofstetter}.
 Comparison between experimental STS
results obtained at different distance values from the impurity
\cite{Madhavan} and theoretical investigations of single impurities
influence on tunneling conductivity \cite{Mantsevich} provide
information weather electron transport occurs coherently or
incoherently and gives an opportunity to initialize impurity type.
All these effects are caused by local changes of the initial density
of states due to interactions of non-equilibrium particles in the
contact area. Taking into account Coulomb interaction of conduction
electrons with non-equilibrium localized charges can result in
nontrivial behavior of tunneling characteristics calculated in the
case of STM metallic tip positioned above the impurity atom
\cite{Arseev,Arseyev,Hofstetter}. However influence of Coulomb
interaction effects on local tunneling conductivity measured apart
from the impurity is a problem of great interest due to the
possibility of impurities types initialization.

    So in this work we present the modification of formula, which describe spatial distribution of local
tunneling conductivity in vicinity of impurity in the case of
interference between resonant and direct tunneling channels
\cite{Mantsevich} due to Coulomb interaction. We performed
calculations for two extreme cases when resonant tunneling takes
place through deep impurity state and through shallow impurity
state. We applied mean field self-consistent approach for shallow
impurity state and in the case of deep impurity non-equilibrium
Coulomb interaction effects were studied with the use of Hubbard-{I}
model. Both approaches were analysed with the use of Keldysh
formalism \cite{Keldysh}. We have found that taking into account
Coulomb interaction in addition to interference between tunneling
channels leads to drastical changing of the tunneling conductivity
form depending on the values of tunneling rates and distance from
impurity.

\section{The suggested model and main results}
\begin{figure}[h]
\centering
\includegraphics[width=70mm]{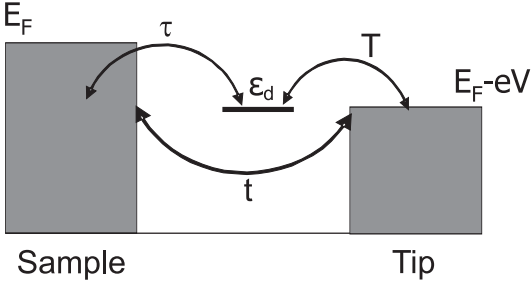}%
\caption {Schematic diagram of direct and resonant tunneling
channels.} \label{1}
\end{figure}

 We shall analyze tunneling between semiconductor surface ($1D$ atomic chain)
and metallic STM tip for deep and shallow impurities in the presence
of Coulomb interaction. The model of tunneling contact formed by
semiconductor and metallic tip is depicted in Fig. \ref{1}. $1D$
atomic chain consists of similar atoms with energy levels
$\varepsilon_{1}$ and similar tunneling transfer amplitudes $\Im$
between the atoms. Distance between the atoms in the chain is the
same and equal to $a$. Atomic chain includes impurity with energy
$\varepsilon_{d}$, tunneling transitions from the impurity atom to
the semiconductor and metallic tip are described by the tunneling
transfer amplitudes $\tau$ and $T$ correspondingly. Direct tunneling
between the surface continuous spectrum states and tip states is
described by the transfer amplitude $t$.

The model system semiconductor-impurity state-metallic tip in the
presence of Coulomb interaction can be described by the Hamiltonian:
$\Hat{H}$:

$$\Hat{H}=\Hat{H}_{0}+\Hat{H}_{imp}+\Hat{H}_{tun}+\Hat{H}_{tip}$$
\begin{eqnarray}
&\Hat{H}_{0}&=\sum_{k\sigma}\varepsilon_{k}c_{k\sigma}^{+}c_{k\sigma}+\sum_{kk^{'}\sigma}\Im c_{k\sigma}^{+}c_{k^{'}\sigma}+h.c.\nonumber\\
&\Hat{H}_{tun}&=\sum_{k}\tau c_{k\sigma}^{+}c_{d\sigma}+\sum_{p}T c_{d\sigma}^{+}c_{p\sigma}+\sum_{kp}tc_{k\sigma}^{+}c_{p\sigma}+h.c.\nonumber\\
&\Hat{H}_{imp}&=\sum_{d\sigma}\varepsilon_{d}c_{d\sigma}^{+}c_{d\sigma}+Un_{d\sigma}n_{d-\sigma};\Hat{H}_{tip}=\sum_{p\sigma}\varepsilon_{p}c_{p\sigma}^{+}c_{p\sigma}\
\label{expression}
\end{eqnarray}
Indices $k$ and $p$ label the states in the left (semiconductor) and
right (tip) lead, respectively. The index $d$ indicates that
impurity electron operator is involved. $U$ is the on-site Coulomb
repulsion of localized electrons,
$n_{d\sigma}=c_{d\sigma}^{+}c_{d\sigma}$, $c_{d\sigma}$ destroys
impurity electron with spin $\sigma$. $\Hat{H}_{0}$ is a typical
Hamiltonian for atomic chain with hoppings without any impurities.
$\Hat{H}_{tun}$ describes resonant tunneling transitions from the
impurity state to the semiconductor and metallic tip and direct
transitions between the tunneling contact leads. $\Hat{H}_{imp}$
corresponds to the electrons in the localized state formed by the
impurity atom, $\Hat{H}_{tip}$ describes conduction electrons in the
metallic tip.

    With the use of diagram technique for non-equilibrium processes \cite{Keldysh}
one can get expression for the spatial distribution of local
tunneling conductivity without Coulomb interaction
\cite{Mantsevich}:

\begin{eqnarray}
\frac{dI}{dV}(\omega,x)&=&\sqrt{\gamma_{kp}\gamma_{kd}\gamma_{pd}\nu^{0}_{k}}ReG^{R}_{dd}(\omega)cos(2k_{x}(\omega)x)+\nonumber\\
&+&\gamma_{kp}(\gamma_{kd}+\gamma_{pd})\nu^{0}_{k}ImG^{R}_{dd}(\omega)cos(2k_{x}(\omega)x)+\nonumber\\
&+&\frac{\gamma_{kd}\gamma_{pd}}{\gamma_{kd}+\gamma_{pd}}ImG^{R}_{dd}(\omega)+\nonumber\\
&+&\frac{\gamma_{kd}^{2}\gamma_{pd}\gamma_{kp}(\omega-\varepsilon_{d})cos(2k_{x}(\omega)x)}{((\omega-\varepsilon_{d})^{2}+(\gamma_{kd}+\gamma_{pd})^{2})^{2}}+\nonumber\\
&+&\gamma_{kp}\nu^{0}_{k}(1+\frac{(\omega-\varepsilon_{d})^{2}+\gamma_{kd}^{2}-\gamma_{pd}(\omega-\varepsilon_{d})}{(\omega-\varepsilon_{d})^{2}+(\gamma_{kd}+\gamma_{pd})^{2}})\cdot\nonumber\\
&\cdot&(\frac{(\omega-\varepsilon_{d})^{2}+(\gamma_{kd}+\gamma_{pd})^{2}(1-cos(2k_{x}(\omega)x))}{(\omega-\varepsilon_{d})^{2}+(\gamma_{kd}+\gamma_{pd})^{2}}+\nonumber\\
&+&\frac{\gamma_{pd}(\gamma_{pd}+\gamma_{kd})cos(2k_{x}(\omega)x)}{(\omega-\varepsilon_{d})^{2}+(\gamma_{kd}+\gamma_{pd})^{2}})\
\label{expression_1}
\end{eqnarray}
where impurity retarded Green's function is defined by the
expression:
\begin{eqnarray}
G^{R}_{dd}(\omega)=\frac{1}{\omega-\varepsilon_{d}-i(\gamma_{kd}+\gamma_{pd})}
\label{expression_2}
\end{eqnarray}
Relaxation rates $\gamma_{kd}$, $\gamma_{pd}$  are determined by
electron tunneling transitions from localized states to the leads
$k$ and $p$ continuum states and relaxation rate $\gamma_{kp}$
corresponds to direct tunneling transitions between $k$ and $p$
continuum states $\sum_{p}T^{2}ImG_{pp}^{0R}=\gamma_{pd}$;
$\sum_{p}t^{2}ImG_{pp}^{0R}=\gamma_{kp}$;
$\sum_{k}\tau^{2}ImG_{kk}^{0R}=\gamma_{kd}$.

$\nu^{0}_{k}$ is unperturbed density of states in semiconductor.
Expression for $k_{x}(\omega)$ can be found from the $1D$ atomic
chain dispersion law:
\begin{eqnarray}
\omega(k_{x})=2\Im\cdot\cos(k_{x}a)\
\end{eqnarray}

\begin{figure*}
\centering
\includegraphics[width=160mm]{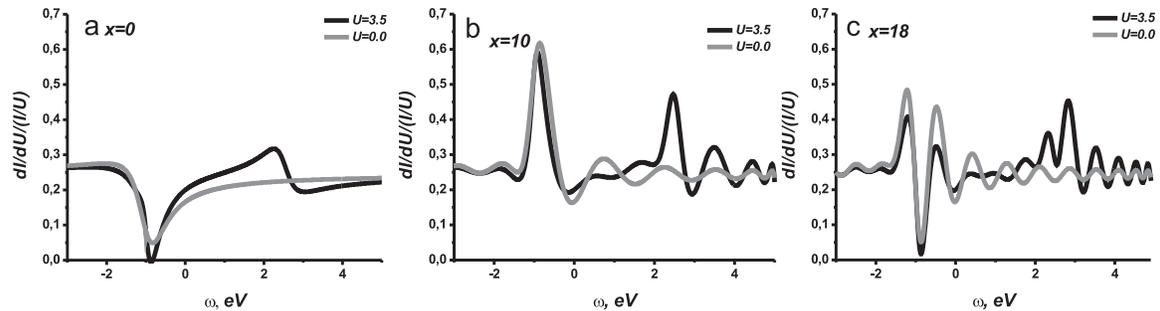}%
\caption{Local tunneling conductivity as a function of applied bias
voltage calculated for different values of the distance $x$ from the
deep impurity atom along the atomic chain in the absence (gray line)
and in the presence (black line) of Coulomb interaction $U$. For all
the figures values of the parameters $a=1$, $t=0,22$, $T=0,18$,
$\tau=0,50$, $\Im=1,00$, $\varepsilon_{d}=-1,00$, $U=3,50$,
$\nu^{0}_{k}=1$ are the same.} \label{2}
\end{figure*}

Let us now analyze modification of formula \ref{expression_1} due to
Coulomb interaction effects in the case of deep impurities with the
help of  Hubbard-{I} model (Hamiltonian of suggested model has the
form of expression \ref{expression} ). Taking into account Coulomb
interaction in this model leads to formation of two well separeted
impurity energy levels $\varepsilon_d$ and $\varepsilon_d+U$ instead
of one initial level $\varepsilon_d$. It is reasonable to use
approximation in which the strongest interaction of the considered
model - the on-site Coulomb repulsion $U$ - is included in
$G^{0R\sigma}_{dd}(\omega)$. So we can write down expression for
$G^{0R\sigma}_{dd}(\omega)$:

\begin{eqnarray}
G^{0R\sigma}_{dd}(\omega)=\frac{1}{\omega-\varepsilon_{d}-\Sigma(\omega)}\nonumber\\
\Sigma(\omega)=\frac{n_{d-\sigma}U(\omega-\varepsilon_{d})}{\omega-\varepsilon_{d}-(1-n_{d-\sigma})U-i\delta}
\end{eqnarray}

With the help of Keldysh diagram technique \cite{Keldysh} using
system of Dyson equations one can get expression for impurity
retarded Green's function.

\begin{eqnarray}
G^{R\sigma}_{dd}=\frac{1}{\omega-\varepsilon_{d}-\Sigma(\omega)-i(\gamma_{kd}+\gamma_{pd})}
\label{expression_6}
\end{eqnarray}

To evaluate the dependence of local tunneling conductivity on the
distance from the deep impurity in the presence of Coulomb
interaction expressions for $ImG^{R}_{dd}(\omega)$ and
$ReG^{R}_{dd}(\omega)$ in formula \ref{expression_1} have to be
retarded by expressions for $ImG^{R\sigma}_{dd}(\omega)$ and
$ReG^{R\sigma}_{dd}(\omega)$. It is also necessary to take into
account the presence of electron spin, it results in extra summation
over $\sigma$ and leads to additional factor equal to $2$ in formula
\ref{expression_1}. Calculating expressions for
$ImG^{R\sigma}_{dd}(\omega)$ and $ReG^{R\sigma}_{dd}(\omega)$
requires solution of self-consistent system of equations. Two of
them are equations for $ImG^{R\pm\sigma}_{dd}(\omega)$, calculated
from expression \ref{expression_6}, and three equations determine
impurity atom non-equilibrium electron filling numbers:

\begin{eqnarray}
n_{d\mp\sigma}=\frac{-1}{\pi}\int d\omega
n_{d\mp\sigma}(\omega)ImG^{R\pm\sigma}_{dd}(\omega,n_{d\pm\sigma})\nonumber\\
n_{d\sigma}(\omega)=n_{d-\sigma}(\omega)=\frac{\gamma_{kd}n^{0}_k(\omega)+\gamma_{pd}n^{0}_p(\omega)}{\gamma_{kd}+\gamma_{pd}}\nonumber\\
\end{eqnarray}

where $n^{0}_k(\omega)$ and $n^{0}_p(\omega)$ are equilibrium
filling numbers in the tunneling contact leads.

\begin{figure*}[t]
\centering
\includegraphics[width=160mm]{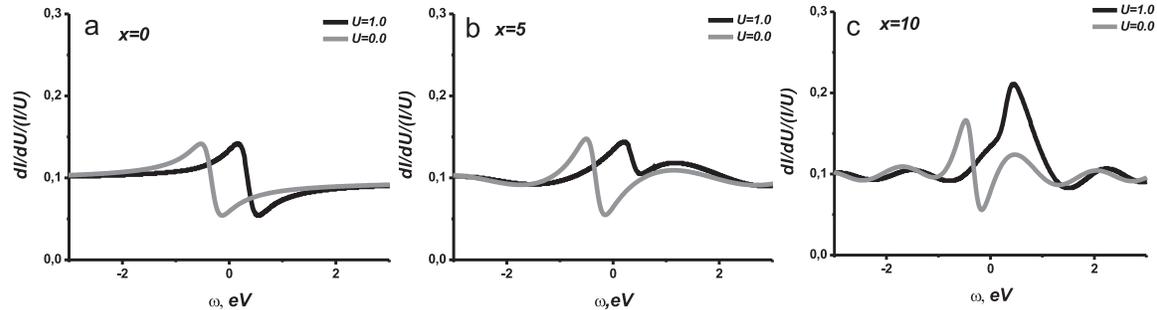}%
\caption{Local tunneling conductivity as a function of applied bias
voltage calculated for different values of the distance $x$ from the
shallow impurity atom along the atomic chain in the absence (gray
line) and in the presence (black line) of Coulomb interaction $U$.
For all the figures values of the parameters $a=1$, $t=0,22$,
$T=0,18$, $\tau=0,40$, $\Im=1,00$, $\varepsilon_{d}=-0,15$,
$U=1,00$, $\nu^{0}_{k}=1$ are the same.} \label{3}
\end{figure*}

Figure \ref{2} shows tunneling conductivity as a function of applied
bias voltage for different values of distance from the deep
impurity. Tunneling conductivity calculated above the impurity has a
resonant dip when applied bias voltage is equal to impurity energy
level position $(\omega=\varepsilon_{d})$ both for taking (Fig.
\ref{2}a black line) and not taking into account Coulomb interaction
effects (Fig. \ref{2}a gray line). Tunneling conductivity calculated
above the impurity in the presence of Coulomb interaction also
reveals peak at the value of applied bias voltage equal to initial
impurity energy level position shifted on the value of Coulomb
potential $(\omega=\varepsilon_{d}+U)$ in comparison with the case
when Coulomb interaction effects are neglected.

At the fixed parameters of tunneling contact existance of a dip or a
peak in tunneling conductivity in both resonances when applied bias
voltage is equal to the impurity energy levels positions is
determined by the value of a distance.

It is clearly evident that in the case of deep impurity state
Coulomb interaction effects lead to negligible shift of a resonant
peculiarity which corresponds to value of applied bias voltage equal
to initial impurity energy level position
$(\omega=\varepsilon_{d})$. Suggested model also demonstrates
absence of resonant peculiarities spreading with increasing of
relaxation rates and distance value from the impurity.

Now let us analyze modification of formula \ref{expression_1} due to
Coulomb interaction effects in the situation of shallow impurity
state. In this case Coulomb interaction effects can be taken into
account with the help of mean-field approximation (adopted
parameters of the model correspond to the mixed valence regime). It
means simply that impurity energy level position depends on Coulomb
interaction of the non-equilibrium electron density. New impurity
energy level position improved by Coulomb interaction can be found
from the equation:

\begin{eqnarray}
\widetilde{\varepsilon}_{d}=\varepsilon_{d}+U\langle n_d\rangle
\label{expression_3}
\end{eqnarray}
where $\varepsilon_{d}$ is the initial position of impurity energy
level without Coulomb interaction. Now the main point is that the
non-equilibrium electron filling numbers $n_d$ for impurity atom
must satisfy self-consistency condition:

\begin{eqnarray}
n_d=\frac{-1}{\pi}\int d\omega n_d(\omega)ImG^{R}_{dd}(\omega)
\label{expression_4}
\end{eqnarray}

where $ImG^{R}_{dd}(\omega)$ is determined from equation
\ref{expression_2} considering $\widetilde{\varepsilon}_{d}$
substitution instead of $\varepsilon_{d}$. Impurity filling numbers
$n_d(\omega)$ can be found from kinetic equations for Keldysh
functions $G^<$ \cite{Mantsevich}:

\begin{eqnarray}
n_d(\omega)=\frac{\gamma_{kd}n^{0}_k(\omega)+\gamma_{pd}n^{0}_p(\omega)}{\gamma_{kd}+\gamma_{pd}}
\label{expression_5}
\end{eqnarray}

 After calculating
impurity atom electron filling numbers $n_d$ from the system of
equations \ref{expression_2}, \ref{expression_3}-\ref{expression_5}
we can determine new position of impurity energy level
$\widetilde{\varepsilon}_{d}$. Next step in  local tunneling
conductivity calculation in the presence of Coulomb interaction for
shallow impurities consists in replacement of energy level value
$\varepsilon_{d}$ by new value $\widetilde{\varepsilon}_{d}$ in
formula \ref{expression_1}.

   Figure \ref{3} shows tunneling conductivity as a function of applied
bias voltage for different values of distance from the shallow
impurity.
%Value of tunneling transfer amplitude from impurity atom
%to metallic tip $\tau$ exceed values of tunneling transfer
%amplitudes from the sample to impurity atom $T$ and to STM tip $t$.
In this case tunneling conductivity calculated above the impurity
has Fano line shape due to interference between resonant and
non-resonant tunneling channels both for taking (black line) and not
taking (gray line) into account Coulomb interaction effects (Fig.
\ref{3}a). Tunneling conductivity calculated without Coulomb
interaction (Fig. \ref{3}a gray line) shows a resonant dip when
applied bias voltage is equal to impurity energy level position
$(\omega=\varepsilon_{d})$. Considering Coulomb interaction for
shallow impurities in the mean field approximation leads to a
resonant dip shift to the higher values of applied bias voltage
(Fig. \ref{3}a black line). This effect also takes place with
increasing of the distance value from the impurity moreover shift is
accompanied by dip's spreading (Fig. \ref{3}b black line). Dip's
spreading is a result of both impurity initial energy level movement
due to Coulomb interaction effects and increasing of relaxation
rates in the studied system. It was found that dip's spreading is
mostly significant when relaxation rate value exceeds value of
initial impurity energy level position $\varepsilon_{d}$. Further
increasing of distance (Fig. \ref{3}c black line) shows disappearing
of the resonant dip in the case of Coulomb interaction opposing to
the calculations carried out without taking into account Coulomb
interaction effects (Fig. \ref{3}c gray line). Disappearing of a dip
is a result of nearest to the dip peaks spreading (one of the peaks
corresponds to the lower value of applied bias voltage and another
to the higher value of applied bias voltage in comparison with dip
position).

\section{Conclusion}
 In this work we have analyzed the role of interplay effects between interference
in tunneling processes and Coulomb interaction in spatial
distribution of tunneling conductivity. We have studied two extreme
cases when resonant tunneling takes place through deep impurity
state and through shallow impurity state. In the case of deep
impurity state non-equilibrium Coulomb interaction effects were
analyzed with the help of Hubbard-{I} model. For shallow impurity
state Coulomb interaction of localized electrons was treated
self-consistently in the mean field approximation.

For shallow impurity state taking into account Coulomb interaction
leads to the shift of a resonant dip to higher values of applied
bias voltage and to the dip's spreading with increasing of tunneling
relaxation rates values in comparison with the case of Coulomb
interaction neglecting.

For deep impurity state taking into account Coulomb interaction
results in formation of additional peculiarity which corresponds to
the value of applied bias voltage equal to initial impurity energy
level position shifted on the value of Coulomb potential.

This work was  supported by RFBR grants and by the National Grants
for technical regulation and metrology.

%\onecolumn

\pagebreak

\end{document}